\documentclass[11pt] {JHEP3}

\usepackage{amsmath,amssymb,amsfonts}
\usepackage[dvips]{graphicx}
 \usepackage{youngtab}
\usepackage{epsfig}

\newcommand{\be}{\begin{equation}}
\newcommand{\ee}{\end{equation}}
\newcommand{\ba}{\begin{eqnarray}}
\newcommand{\ea}{\end{eqnarray}}
\newcommand{\bea}{\begin{array}}
\newcommand{\eea}{\end{array}}

\newcommand{\tr}{{\rm tr}}

\makeatletter \@addtoreset{equation}{section} \makeatother

\preprint{KIAS-P08028 \\
Imperial/TP/08/SK/02}

\title{Dyonic Instantons \\
 in 5-dim Yang-Mills Chern-Simons Theories}

\author{Seok Kim \\
Theoretical Physics Group, Blackett Laboratory, \\
Imperial College, London SW7 2AZ, U.K.  \\
\email{s.kim@imperial.ac.uk }  }

\author{Ki-Myeong Lee \\
School of Physics \\
Korea Institute for Advanced Study \\
Seoul 130-012, KOREA\\
\email{klee@kias.re.kr} }

\author{Sungjay   Lee\\
School of Physics \\
Korea Institute for Advanced Study \\
Seoul 130-012, KOREA\\
\email{sjlee@kias.re.kr} }

\abstract{We consider the BPS dyonic instantons in the 5-dim supersymmetric Yang-Mills Chern-Simons theories.
Its field theoretic structures and the moduli space dynamics in term of the ADHM data have been explored in
detail. We find that   the field theoretic Chern-Simons term  leads to an effective magnetic field on  instanton
moduli space.   }

\begin{document}

\section{Introduction and Conclusion}

 %%%%%%introduction

Dyonic instantons in 5-dim Yang-Mills theories in the Coulomb phase have been explored  in many  different point
of view. Electric Coulomb repulsion balances the force to shrink instantons in the Coulomb phase~\cite{LT99,Zamaklar:2000tc,Eyras:2000dg}. Dyonic  instantons  in the Coulomb phase represents a typical example of supertubes realized in a field theory.  In the Coulomb phase, it has  been argued that dyonic instantons has magnetic monostrings imbedded interior
along which distributed are   instanton  electric charge and linear  momentum~
\cite{Bak:2002ke,KL03,Kim:2006ee,CEH06,CKLL07}. Recently  dyonic instantons configurations are studied  in 5-dim ${\cal N}=1$ supergravity model with nonabelian
gauge group~\cite{KL07}.  A new feature is the 5-dim Chern-Simons term in the action, which leads
  nonabelian electric charge on   instantons even in the symmetric phase without costing additional energy.

%%%% the goal of this work
In this work we investigate  dyonic instantons in the 5-dim ${\cal N}=1$ supersymmetric Yang-Mills theories
with Chern-Simons term.  While the kinetic energy is not positive definite, we were able to characterize the 1/2
BPS supersymmetric dyonic instanton  configurations which saturate  a `pseudo' BPS energy bound.  The low
energy dynamics of these dyonic instantons is described by the  gauged matrix mechanics theory
corrected by a simple background charge term, which reduces the number of  supersymmetries to four. The
moduli space dynamics of these dyonic instantons is a modification of the standard instanton moduli space
dynamics by a term which is first order in time derivative and so can be attributed as  a background magnetic field
on the moduli space. We work out several examples to illuminate the various aspects of the dyonic instantons
and also the quantum aspect of nonabelian charge for a single instanton in the symmetric phase.

 %%% background material

 The 16 supersymmetric 5-dim $U(N)$ Yang-Mills theories can arises from the low energy dynamics of $D4$  branes.  While these theories are not renormalizable, there exists an ultraviolet completion in terms of the field theory,  the so-called $(2,0)$ theory compactified on a circle of radius $g^2/4\pi$ with the gauge coupling constant $g$, and the instantons play  the KK mode  on the circle. In the Coulomb phase, the instantons collapses. However, there exist regular dyonic instanton configurations in the Coulomb phase, which can be regarded as composite of F1 and D0 branes on D4 branes. In the Coulomb phase, there is a potential induced on the moduli space dynamics without breaking the supersymmetry~\cite{LT99}.  There is a natural induced D2 brane dipole charge which can be regarded as the magnetic monostrings inside the dyonic instantons, and so these dyonic instantons in the Coulomb phase is a typical example of a supertube.  These dyonic instantons have been studied in detail~\cite{Bak:2002ke,KL03,Kim:2006ee,CEH06,CKLL07}.

The 8  supersymmetric  5-dim $U(N)$ Yang-Mills theories
with flavor hypermultiplets have been approached in somewhat
differently as their brane counterparts are more complicated. When
the flavor number is small, there can be a ultraviolet  fixed
point with enhanced global symmetries. The Chern-Simons term can
be also induced by    the massive hypermultiplet in the symmetric phase
 and by the massive gauginos and hypermultiplets in the Coulomb phase~\cite{Seiberg96}. The kinetic energy of the  classical theory with the Chern-Simons term is not positive definite anymore  when the scalar field is large. But one expects there is an ultraviolet complete
theory for a small flavor case, for which the kinetic energy
remains nonnegative~\cite{Seiberg96,MS96}. A complete classification
of all possible nontrivial five-dimensional fixed points of
gauge-theoretic origin as been give in Ref.~\cite{IMS97}.

Such 8 supersymmetric 5-dim Yang-Mills theories can appear in the
brane dynamics as low energy dynamics of the D5- NS5
web dynamics~\cite{AH97,AHK97}. The additional massive fundamental hypermultiplets
can be introduced either  by spectator D7 branes or D5
branes. The integration of the massive hypermultiplets would lead
to the Chern-Simons term. The instantons in 5-dim gauge theories with Chern-Simons term
have been discussed briefly in relation to skyrmions in Ref.~\cite{Eto:2005cc,Hong:2007ay}.

%%%%What we have done here.

 The 5-dim Chern-Simons term is possible for $SU(N)$ gauge theories  with $N\ge 3$, and for those  gauge theories with $U(1)$ factor.  The Chern-Simons term can be induced by the massive fermions and  its coefficient $\kappa$   should be quantized for the consistency of the path integral. On noncommutative four-space,  the natural gauge group is $U(N)$ and the Chern-Simons term is possible for all $N$. The instantons   cannot collapse on noncommutative space, and the low energy dynamics described by the gauged sigma model is modified by the FI term, which makes the instanton moduli space smooth. While our primary focus is on dyonic instantons on commutative space, we will make some discussion for the noncommutative case also. (See ref.~\cite{Douglas:2001ba} for a review on the
noncommutative field theory.)

The 1/2 BPS dyonic instanton configuration is characterized by   (anti-)selfdual instanton configurations. In the symmetric phase, the Gauss law  leads to  the electric   field profile for a given instanton configuration. For a single instanton configuration in the $SU(N)$ gauge theory we find the scalar field profile explicitly which is singular when the instanton size is less than of order $\kappa g^2$. This shows that the ultraviolet cutoff length scale of the Yang-Mills theory with the Chern-Simons term is somewhat larger than the natural scale $g^2$ for large $\kappa$. We do not know the stringy origin of this gap.

The low energy dynamics of the $K$ number of instantons for the 16
supersymmetric theories have 8 supersymmetric gauged sigma model
with hypermultiplets in the adjoint and fundamental representation
of $U(K)$  gauge symmetry. The noncommutative space can be taken
into account by the F-I term in this gauge theory without breaking
any supersymmetry. We argue that the 5-dim Chern-Simons term in
the field theory leads to a 1-dim Chern-Simons term
in the gauged linear sigma model, which is  basically an
interaction term between the $U(K)$ gauge field and
the background charge. The coefficients of two Chern-Simons  terms are
identical up to an obvious multiplicative factor. This 1-dim
Chern-Simons term also reduces the supersymmetry of the gauged
sigma model by half. While one needs an ultraviolet
completion of the 5-dim theory, the 1-dim gauged sigma model for
dyonic instantons is complete by its own. This leads to a tiny
gap for the electric charge energy calculated in the field theory
and the gauged sigma model in the Coulomb phase.

One can reduce further the gauged sigma model to nonlinear sigma
model on the instanton moduli space by solving the vacuum equation
of this sigma model and the gauge field equation. The 1-dim Chern-Simons term leads to a
background magnetic field on the instanton moduli space. (This induced magnetic field was also noted   in the context of the Nekrasov's instanton counting with 5-dimensional Chern-Simons terms \cite{tachikawa}.) We find this magnetic field on the instanton moduli space explicitly for a single instanton on commutative and non-commutative spaces.  For a single instanton
on the commutative space, we quantize the moduli space dynamics
and show what kind of the `nonabelian' electric
charge a single instanton   can carry   in the symmetric phase.

%%%%%%%%% Conclusion
Our theory allows also 1/2 BPS magnetic monostrings in the Coulomb phase. As in the maximally
supersymmetric case, instantons and W-bosons are attracted to monostrings.  In addition, one can get the linear momentum  be induced on the monostrings.  Similar to Ref.~\cite{Kim:2006ee}, we can easily obtain the BPS energy bound on   magnetic monostrings with both instanton and electric charge density. This shows again the supertube origin of our dyonic instantons in the Coulomb phase.  There is some investigation of the magnetic monostrings in the context of anomaly~\cite{Boyarsky:2002ck}.  Our work should shed some further light on the subject.

There are a couple of directions one can explore starting from our work.  The role of the instanton solitons  in 5-dim Yang-Mills  theories is not fully explored.  In the Coulomb phase,  there is a 1-loop correction to the Chern-Simons term   due to the massive gauginos and hypermultiplets.  This would lead to quantum modification of the characteristics   of the dyonic instantons.

 When one of the spatial direction is compact, instantons or calorons  can be decomposed into magnetic  monopoles.  Our dyonic instantons would be also decomposed into dyons. When the hypermultiplets are present and  when the theory is in the Higgs phase, one could have the magnetic flux vortices and magnetic monopoles   would be in trapped between flux vortices.~\cite{Tong:2005un,Eto:2004rz} We expect similar composite objects in the theory with Chern-Simons term.

  %%%%%%% Outline

  The plan of this paper is as follows. In Sec. 2,  we review the basic properties of the 5-dim ${\cal N}=1$ supersymmetric Yang-Mills theories with the Chern-Simons term. In Sec. 3, we consider the 1/2 BPS dyonic
  instanton configurations from the energy bound point of view and solve the Gauss law explicitly
  for a single instanton in $U(N)$ theory. In Sec.4, we discuss the gauged sigma model with 1-dim Chern-Simons
  term and also the moduli space dynamics with a background magnetic field. In Appendix, we review
  some aspects of the ADHM formalism for instantons.

\noindent{\bf Note Added:} After this work has appeared in arxiv, a work of some  overlap  appeared ~\cite{Collie:2008vc}. 
 
\section{5-dim ${\cal N}=1$  Supersymmetric  Gauge Theories}

\subsection{the Lagrangian}

Let us fix the Lagrangian for 5-dim $ {\cal N}=1$ supersymmetric gauge  theories with  the 5-dim
Chern-Simons term by using the 4-dim ${\cal N}=2$ supersymmetric Yang-Mills theories.
To find the supersymmetric counterpart for the Chern-Simons term, we consider the $U(1)$ theory on a circle. The
low energy 4-dim Lagrangian theory would have terms  at most quadratic in the spacetime derivatives. As the 4-
dim theory has the eight supersymmetries, it is decided by the prepotential ${\cal F}(A)$. Especially the $A_5$
plays the
role  of scalar field and the 5-dim Chern-Simons term would be come the 4-dim axion-photon coupling term.
In addition, five-dimensional gauge invariance highly restricts the
prepotential to be at most cubic polynomial; Invariance under $A_5
\rightarrow A_5 + a$ with $a$ arbitrary real constants translates
to invariance under the shift   ${\cal A} \rightarrow {\cal A} + i a$ of
the chiral field ${\cal A}= \phi+ iA_5$.   Assuming the prepotential to have the cubic term ${\cal A}^3$ so that
\be
 {\cal F} \sim  {\cal A}^3
 \ee
 the 4-dim bosonic Lagrangian becomes
 \be
  {\cal L} \sim   -  \phi
 \left(  ( \partial_\mu \phi)^2 + (\partial_\mu A_5)^2  + \frac12 F_{\mu \nu} F^{\mu \nu}
 \right) +  A_5 \left( \frac14 F_{\mu \nu} \tilde{F}^{\mu \nu}
 \right) + \cdots,
\ee
which is invariant under the constant shift $A_5 \rightarrow
A_5 + a$. When the prepotential has terms higher than cubic
order, the Lagrangian is no longer invariant under the constant
shift. For instance, let us consider the quartic prepotential
${\cal F}= \frac{1}{4 !} {\cal A}^4$. The Lagrangian now contains
\begin{eqnarray}
  {\cal L} ~\sim~ \left (\phi^2 - A_5^2 \right) F_{\mu \nu} F^{\mu
  \nu} + \left( 2 \phi A_5 \right) F_{\mu \nu} \tilde{F}^{\mu \nu}
  + \cdots,
\end{eqnarray}
which is not invariant obviously under  the shift $A_5\rightarrow A_5+a $.

We therefore conclude that the most general five dimensional
supersymmetric $SU(N)$ gauge theory can be described by the cubic
prepotential
\ba
  {\cal F}(\mathcal{A})&=&{\rm tr}_N\left(\frac{1}{g^2}\mathcal{A}^2+\frac{\kappa}{3}
  \mathcal{A}^3\right) \nonumber \\
  &=& \frac{1}{2g^2}\mathcal{A}^a\mathcal{A}^a+
  \frac{\kappa}{3}d_{abc}\mathcal{A}^a\mathcal{A}^b\mathcal{A}^c   \ .
\ea
Here we used the convention for the traceless $SU(N)$ generators $T_a$  as
\be
  \{ T_a , T_b \} = \frac1N \delta_{ab} 1_N + 4 d_{abc} T_c ~.
\ee
so that  $\tr T_aT_b=\delta_{ab}/2$ .
The bosonic part of the action is therefore
\begin{eqnarray}
  S&=&\int d^5 x\ {\rm tr}\Big(-\frac{1}{2 g_{\rm eff}^2} \big( F_{\mu\nu}F^{\mu\nu} +2
D_\mu\phi
  D^\mu\phi \big) \Big)+S_{CS}\\
  &=&\int d^5 x\ \Big(-  \frac14 \big(\frac{1}{g^2_{\rm eff}}\big)_{ab} \big( F^a_{\mu\nu}F^{b\mu\nu} +
  2D_\mu\phi^aD^\mu\phi^b \big) \Big)+S_{CS}\nonumber
\end{eqnarray}
with
\ba\label{effgaugecoupling}
  \frac{1}{g_{\rm eff}^2} = \frac{1}{g^2}{\bf 1}_N + \kappa \phi\ , \hspace{1cm}
  \left( \frac{1}{ g_{\rm eff}^2 } \right)_{ab}= \frac{1}{g^2} \delta_{ab} + 2\kappa d_{abc}\phi^c\ .
\ea
The 5-dim Chern-Simons term is defined as
\ba \label{CS}
  S_{CS} &=&  \frac{\kappa}{3}\int{\rm tr}\left(A\wedge F\wedge F+\frac{i}{2}
  A\wedge A\wedge A\wedge F-\frac{1}{10}A\wedge A\wedge A\wedge A\wedge
  A\right)\ \nonumber \\
  &=& \frac{\kappa}{12}\int d^5x \; \epsilon^{\mu\nu\rho\sigma\chi} \Big( d^{abc} A^a_\mu F^b_{\nu\rho}
F^c_{\sigma\chi} + ...\Big)\ ,
\ea
which allows only eight real supersymmetries rather than sixteen. The Gauss law constraint for all physical
configurations   is given by the traceless part
of the following equation
\be \label{Gauss1}
  -  D_m \big\{ \frac{1}{g_{\rm eff}^2} , E_m \big\}
 -i   \Big[\phi, \big\{  \frac{1}{g_{\rm eff}^2} , D_0\phi]\big\} \Big ]
  + \frac{\kappa}{2} F_{mn} \tilde{F}_{mn}  =
  0\ ,
\ee
where  $E_m\equiv F_{m0}$ and the dual of the spatial field strength is
\be \tilde{F}_{mn} \equiv \frac12 \epsilon_{mnpq} F_{pq}\ . \ee

Of course the 5-dim Yang-Mills theory is not renormalizable as the coupling constant $g^2$ has the dimension of
the length. We thus consider the field theory with cut-off while it is possible to imagine a consistent quantum
completion of the theory such that additional degrees of freedom appears in the ultraviolet region in the energy
scale larger than $1/g^2$. With the Chern-Simons term, another length  scale $ \kappa g^2 > g^2$ could be
introduced.  Later we will see the classical instanton configuration breaks down in some cases when the
instanton size is less than the order of  $\kappa g^2$.

The Chern-Simons coefficient is quantized for the consistency of the path integral. Note that the Chern-Simons
term vanishes for the $SU(2)$ gauge group as there is no nontrivial $d_{abc}$ coefficient.
When one makes  a large gauge transformation with nontrivial homotopy as $\pi_5(SU(N))={\bf Z}$ for $N\ge
3$,
the Chern-Simons term transforms so that the topological quantity for this homotopy group appears.
The quantum consistency requires the quantization,
\begin{eqnarray}
  \kappa ~=~ \frac{n}{8\pi^2} \ , \ \ \ \left( n \in  {\bf Z} \right)\ ,
\end{eqnarray}
where we call the integer quantity $n$ as the {\it Chern-Simons level}.

One important feature of the Chern-Simons theories is that the Chern-Simons couplings can be induced by the
radiative  quantum corrections by a massive Dirac fermion\cite{Witten96}. Let us consider an amplitude ${\cal A}$
involving three external non-abelian gauge bosons of momentum $p$, $q$ and $-(q+p)$ together with gauge and
Lorentz indices $(a,\mu)$, $(b, \nu)$ and $(c, \rho)$, respectively. The parity violating term can be expressed as
\begin{eqnarray}
  {\cal A}_{\mu\nu\rho}^{abc} &=& - \int \frac{d^5k}{(2\pi)^5}\  \text{tr} \left( \gamma_{\mu} \frac{1}{k\cdot \gamma
+im} \gamma_{\nu}
\frac{1}{(k-q)\cdot \gamma +im} \gamma_{\rho} \frac{1}{(k+p)\cdot \gamma +im}\right)\,  \text{tr}\left( T^a \{ T^b,
T^c\} \right) \nonumber \\
  &=& \frac{i}{16\pi^2} \frac{m}{|m|} \epsilon_{\mu \nu \rho \gamma \delta} q^\gamma p^\delta  \text{tr}\left(T^a
\{ T^b, T^c \} \right),
\end{eqnarray}
where the gamma matrices $\gamma_{\mu}$ here should satisfy the Dirac-Clifford algebra relation
 $\{ \gamma_{\mu} , \gamma_{\nu} \}=  2\eta_{\mu \nu}$ with $\gamma_{01234}=i$. One can therefore obtain
the induced couplings after integrating out the $N_f$ massive hypermultiplets
\begin{eqnarray}
  {\cal L}_{\text{induced}}&=&  \text{sign}(m) \frac{N_f}{48 \pi^2}  \int d^5x \ \epsilon^{\mu \nu \rho \gamma
\delta} A_\mu^a \partial_{\gamma} A_{\nu}^b \partial_{\delta} A_\rho^c \,
  \text{tr} \left( T^a T^b T^c \right)\nonumber \\ &=&
  -  \text{sign}(m) \frac{N_f}{2} \frac{1}{24 \pi^2} \int~ \text{tr} \left( A \wedge F \wedge F \right)\ .
\end{eqnarray}
Taking account of four and five external gauge bosons diagram in additions, one can find the induced shift of the
Chern-Simons level
\be
\Delta n = -\frac{N_f}{2} \ee
If the zeroth order Chern-Simons term vanishes,  one requires the even numbers
of hypermultiplets for the gauge invariance. It agrees with the
result in \cite{IMS97}.

When the gauge group is $U(N)$, the gauge coupling constant for $U(1)$ theory and that for the $SU(N)$ theory
can be different in general.  There could be also  Chern-Simons terms for the purely $U(1) $ theory and the mixed
term between $SU(N)$ and the $U(1)$ parts with different coefficients. On noncommutative 4-dim space, the
gauge group however should be $U(N)$ instead of the $SU(N)$, and so there exist only one gauge coupling
constant and also only one Chern-Simons
coefficient. It is noteworthy here that, when we consider the noncommutative five-dimensional $U(N)$ gauge
theory, all traceless condition for $SU(N)$ gauge symmetry should be dropped out.

\subsection{Brane Configuration}

In this section, we will present certain brane configurations describing the  five-dimensional ${\cal N}=1$
supersymmetric gauge theory \cite{AH97,AHK97} with the Chern-Simons couplings. These configurations leads
later a useful way to consider the low energy dynamics of   dyonic instantons in terms of the dynamics of the D-
branes corresponding to instantons.

Let us start with two parallel solitonic (NS) fivebranes, $N$ D5-branes suspended between two NS5-branes.
Their configurations are summarized in the table \ref{brane}. It is noteworthy here that the brane configuration
above preserves quarter of maximal supersymmetries, that is, eight real supersymmetries. Since D5-branes are
finite in $x^6$ direction, their macroscopic description is five-dimensional ${\cal N}=1$ $G=SU(N)$ super Yang-
Mills theories. The five-dimensional gauge coupling can be now described as
\begin{eqnarray}
  \frac{1}{g^2}~=~\frac{\Delta x^6}{(4\pi g_s) (2 \pi l_s)^2 } \hspace{0.5cm} \rightarrow \hspace{0.5cm}
\left[\frac{1}{g^2}\right]~=~\text{mass}\ ,
\end{eqnarray}
where $l_s$ and $g_s$ are the string length and coupling respectively. When D5 branes meets NS5 branes, they
form a (p,q) web of 5-branes, and so the $U(1)$ group factor which is responsible for the motion of the center of
the mass of D5 brane segments are frozen as discussed in   \cite{Witten97}.  It would be interesting find out why
this is not case when the noncommutativity on D5 branes are introduced.

For the field theory decoupling limit, we should require the physical states in the field theory not to excite the
Kaluza-Klein modes and stringy states. In other words, one must require the characteristic energy scale $\frac{1}
{g^2}$ in the field theory to be smaller than
\begin{eqnarray}
  \frac{1}{g^2}~\ll~ \frac{1}{l_s}\ , \frac{1}{\Delta x^6}\ ,
\end{eqnarray}
while taking the $g_s \rightarrow 0$. One can therefore conclude that the decoupling limit is given by
\begin{eqnarray}
  \Delta x^6 ~\ll~ g_s l_s\ .
\end{eqnarray}
When NS5-branes are far apart than $l_s$, one can recover six-dimensional theories with various stringy effects.
\begin{table}
$$
\begin{array}{l|ccccc|ccccc}
  \hline \hspace{0.3cm}
  &\hspace{0.2cm} 0 \hspace{0.2cm}&\hspace{0.2cm} 1 \hspace{0.2cm}&
  \hspace{0.2cm}2\hspace{0.2cm} & \hspace{0.2cm}3\hspace{0.2cm} & \hspace{0.2cm}4\hspace{0.2cm} &
\hspace{0.2cm}5\hspace{0.2cm} & \hspace{0.2cm}6\hspace{0.2cm} & \hspace{0.2cm}7\hspace{0.2cm} &
\hspace{0.2cm}8\hspace{0.2cm} & \hspace{0.2cm}9\hspace{0.2cm}  \\
  \hline
  \text{D5} & \circ & \circ & \circ & \circ & \circ & & \circ & & & \\
  \text{NS5} & \circ & \circ & \circ & \circ & \circ & \circ & & & &\\
  \text{D7} & \circ & \circ & \circ & \circ &\circ  &  & & \circ & \circ & \circ\\
  \hline
\end{array}
$$
\caption{Brane configuration for 5-dimensional ${\cal N}=1$ supersymmetric gauge theories}\label{brane}
\end{table}

In five-dimensional gauge theories, there is an additional $U(1)_I$ symmetry of interests, whose conserved
current takes the form
\begin{eqnarray}
  j ~=~ \frac{1}{8 \pi^2} \ast_5 \text{tr}\left( F\wedge F\right).
\end{eqnarray}
Its charge is the instanton number. In our brane picture, an instanton whose $U(1)_I$ charge is unit can be
identified as a D1-brane stretched along the $x^6$ direction. It is depicted in Fig. One simple way to check this
identification follows from the agreement in the mass
\begin{eqnarray}
  \frac{8\pi^2}{g^2}~=~\frac{\Delta x^6}{g_s (2\pi l_s)}~=~\tau_{D1}\Delta x^6\ ,
\end{eqnarray}
where $\tau_{D1}$ denotes the tension of a D1-brane.

In order to introduce massive $N_f$ hypermultiplets, there are two options, of which (1) one is to add the $N_f$
D7-branes whose world volume is parameterized by $01234789$ coordinates and (2) another is to add $N_f$
flavor semi-infinite D5-branes. Their configurations are also summarized in the table \ref{brane}, and depicted in
Fig.1.  As discussed in \cite{HW96}, the open string modes connecting D5 with D7 branes for the case (1), or D5
with flavor fivebranes for the case (2) give rise to massive hypermultiplets of masses $m=\frac{\delta x}{2\pi l_s^2}$.   While introducing D7-branes changes the global property of the space-time geometry, which requires a
more careful analysis in general, we ignore it here. For our discussion, either picture is fine\cite{AH97}.

The Chern-Simons coupling can be finally introduced after integrating out these massive fermions in the
hypermultiplets.  From D5 brane point of the view, it is exactly the induced Chern-Simons term by the massive
Dirac fermions as discussed in the previous subsection.

 \begin{figure}[htb]
 \begin{center}
 \includegraphics[width=10cm]{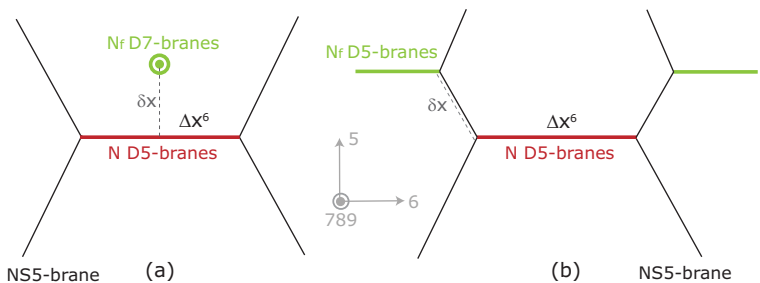}
 \caption{Brane configurations for 5d supersymmetric gauge theory} \label{brane1}
 \end{center}
 \end{figure}

As a final comment, let us  consider the Coulomb branch of the theories in the presence of the Chern-Simons
couplings. In our brane setting, the Chern-Simons terms are in fact induced by integrating out the massive
hypermultiplets. When we try to analyze the Coulomb branch of theory with those brane configurations, we
therefore have to suppose that Higgs vevs, or distance between several D5-branes are much smaller than the
masses of hypermultiplets or $\delta x$.

\subsection{Vacuum Moduli}

We now in turn discuss the Coulomb branch of the present theories and non-trivial fixed points. In five-
dimensional ${\cal N}=1$ supersymmetric gauge theories, the Coulomb branch is parameterized by real scalar
fields of vector multiplets. For the gauge group $SU(N)$ of rank $N-1$, the gauge symmetry is spontaneously
broken down to $U(1)^r$ at a generic point on the moduli space. Weyl reflection symmetries implies that the
Coulomb branch can therefore be identified with  a Weyl chamber ${\bf R}^{N-1}/{\cal W}$.

We can naively expect from the gauge coupling of negative mass dimension that the low-energy effective theories
on the Coulomb branch are trivial. Seiberg has however shown in \cite{Seiberg96} that under one-loop corrections
there exist certain strongly-coupled non-trivial fixed point. Although those topics are not relevant in our discussion
later, basic ideas will be presented briefly.

For simplicity, let us consider the $SU(2)$ gauge theory with $N_f$ hypermultiplets of masses $m_i$
($i=1,2,\cdots,N_f$). For a generic value of $\phi$, a scalar field of vector multiplet, the induced Chern-Simons
level is proportional to
\begin{eqnarray}
  n_{\text{induced}} \ = \  2\, \text{sign}(\phi) -  \frac18  \sum_{i=1}^{N_f} \Big[ \text{sign}
(\phi - m_i) + \text{sign}(\phi + m_i) \Big]\ ,
\end{eqnarray}
where the former contribution is from massive gauginos and the latter from massive hypermultiplets. This implies
that the effective gauge coupling (\ref{effgaugecoupling}), or metric on the Coulomb branch, takes the form as
\begin{eqnarray}
  \frac{1}{g_{\text{eff}}^2} ~=~ \frac{1}{g^2} +  \frac{1}{16\pi^2} \left( 2|\phi| - \frac{1}{8}\sum_{i=1}^{N_f} \Big[ |
\phi-m_i| +  |\phi + m_i | \Big] \right)\ ,
\end{eqnarray}
after finite constant shift of $\frac{1}{g^2}$. We will assume from now on that all mass parameters are turned off.

This metric implies that for $N_f > 8$, there is a singularity at a finite point $\phi \sim \frac{1}{g^2}$ in the moduli
space, reflecting the non-renormalizability of the present model. As discussed in \cite{Seiberg96}, non-existence
of the singular points on the Coulomb branch for $N_f < 8$ implies the non-trivial fixed point together with the
enhancement of global symmetries in the strong coupling limit $g \rightarrow \infty$. Possibility of those non-trivial
RG fixed points for various models with higher rank gauge groups and the classification of them in accordance
with the classification of singularities of Calabi-Yau threefolds have been presented extensively in
\cite{MS96,IMS97}.

\section{BPS Configurations}

\subsection{BPS Energy Bound and BPS Equations}

We now in turn consider the BPS bound for the present model. The energy functional now
takes the form
\begin{eqnarray}
  {\cal E} ~=~ \int d^4{\bf x} \, \tr_N \left( \frac{1}{g_{\rm{eff}}^2}
   \Big[ E_m^2 +\frac{1}{2}F_{mn}^2
  + (D_0\phi)^2 + (D_m\phi)^2 \Big]  \right)~.
\end{eqnarray}
In order to make the energy positive definite, we will consider the family of the field configurations such that
$g_{\rm eff}^2>> 0$. Thus we consider the region where the value of $1/g^2$ is much larger than the value of $
\kappa \phi $ for the sake of argument. We rearrange
the above energy functional as
\ba
  {\cal E} &=& \int d^4{\bf x}\,\tr_N \Big(   \frac{1}{g_{\rm{eff}}^2}   \Big[
  (E_m-\alpha D_m\phi)^2 + \frac{1}{4}
  (F_{mn}-\alpha \tilde{F}_{mn})^2   \nonumber
  \\
  & & \;\;\;\,\,\,\,\,  +(D_0\phi)^2  +\alpha  \{E_m,D_m\phi\} + \frac{\alpha}{4}
  \{F_{mn},\tilde{F}_{mn}\}  \Big] \Big)
  \nonumber \\
  &\geq& \alpha  \int d^4{\bf x}\,\tr_N \Big(  \frac{1}{g_{\rm eff}^2}
   \Big[ \{E_m,D_m\phi\} +
  \frac{1}{4}\{F_{mn},\tilde{F}_{mn}\}
  \Big]\Big)  \nonumber \ea
where $\alpha=\pm 1 $. One can show from the Gauss law (\ref{Gauss1}) that the total energy is bounded by
\be {\cal E} \ge  \Big|  \frac{8\pi^2}{g^2} K + Q_E \Big|\ ,  \ee
where the topological charge and the electric charge energy are defined
respectively as
\be K = \frac{1}{16\pi^2} \int d^4x\,  \tr_N  ( F_{mn}\tilde{F}_{mn}) \,
  , \,\, \;\; Q_E =
\int_{S_{\infty}^3}  dS^m \tr_N  \Big( \phi  \big\{ \frac{1}{g_{\rm eff}^2}
 , E_m \big\}   \Big) \ . \ee
It should be noted here that the energy due to the electric charge
vanishes in the symmetric phase where $\phi_\infty=0$,
even though the instanton can carry nontrivial `nonabelian' electric charge.
The bound is saturated by the BPS configurations satisfying the
Gauss law (\ref{Gauss1}) and the BPS equations
\be F_{mn}-\alpha \tilde{F}_{mn}=0\, , \, E_m-\alpha D_m\phi=0\; ,
\; D_0\phi=0\ .  \label{bps1}\ee

In order to see that these BPS configurations preserve four supersymmetries, let us consider the
supersymmetry transformation rules of the theory.  As discussed in section 2, the theory at hands has eight
supersymmetries whose the supersymmetric parameter satisfies the 6-dimensional Weyl condition
\be  \Gamma^{012345}\epsilon=\epsilon\ , \label{weyl}\ee
together with the 10-dimensional Majorana and Weyl condition. (Here we use the 10-dimensional notation for the
Gamma matrices, and the gaugino field $\lambda$ also satisfies all the same conditions as $\epsilon$.) For the
BPS configurations, the supersymmetric transformation of the gaugino field
\be \delta \lambda = \frac12 \Gamma^{MN}F_{MN}\epsilon\ , \ee
where $A_5=\phi$, is required to vanish. Imposing a supersymmetric condition compatible with the Weyl
condition (\ref{weyl})
\be \Gamma^{1234}\epsilon=\alpha  \epsilon, \ \text{ or, equivalently } \ \Gamma^{05}\epsilon=\alpha \epsilon \ ,\ee
one can recover the BPS equations (\ref{bps1})
\be F_{mn} - \alpha \tilde{F}_{mn} =0 \;   , \; F_{m0} -\alpha D_m
\phi=0\; , \; D_0\phi=0 \ .\label{BPS1} \ee
This implies that the BPS configuration preserves the half of the supersymmetries. Unlike the ${\cal N}=2$
theories, there is a correlation between the sign of the instanton charge and that of electric charge in the ${\cal
N}=1$ theory, regardless the presence of the
Chern-Simons term.

Once we choose the gauge  $A_0= \alpha \phi$, the field configuration becomes static in time and the Gauss law
(\ref{Gauss1}) can be reduced to
\begin{eqnarray}\label{Gauss3}
 - D_m \left\{\frac{1}{g_{\rm eff}^2}, D_m\phi \right\}
  + \frac{\kappa}{2} F_{mn}^2 = 0\ .
\end{eqnarray}
One can further reduce the Eqn. (\ref{Gauss3}) by using the BPS equations as
the traceless part of the following equation,
\be \label{gauss2}  D_m^2 \Phi =  \frac{\kappa}{4} F_{mn}^2 \; ,
\ee
where
\be \Phi \equiv \frac{1}{g^2}\phi+\frac{\kappa}{2} \left( \phi^2 - \frac{\text{tr}(\phi^2)}{N} {\bf 1}_N \right)\ee
or, explicitly
\ba \label{BPS2}
  D_m^2 \Phi^a ~=~ \frac{\kappa}{2} d_{abc}
  F_{mn}^b F^{c}_{mn}\ ,
\ea
where $\Phi=\Phi^a T^a$ and $ \Phi^a =\frac{\partial}{\partial\phi^a}{\cal
F}(\phi)=\frac{1}{g^2}\phi^a+ \kappa d_{abc}\phi^b\phi^c$. This scalar equation does not depend on the choice of
$\alpha$. Here we want to note that BPS equations (\ref{BPS1},\ref{BPS2})
are exactly the same to those obtained from the SUGRA
analysis\cite{KL07}.

Before closing this subsection, we now in turn present the physical charges of our BPS solutions. We have to
define the electric charge more carefully in the present model. We can usually read off the electric charge from the
asymptotic behavior of the displacement vector ${\cal D}_m=\frac{\delta {\cal L}}{\delta \dot{A}_{m}}$
\ba \label{electric1}
  {\cal D}_m &=& \{ \frac{1}{g_{\rm eff}^2} , E_m \} - \text{ (trace part) }~=~ \frac{  x_m}{4\pi^2 r^4} Q_N+ {\cal O}
(\frac{1}{r^4}) ,
\ea
where $Q_N$ is the `nonabelian' electric charge of the system. Assuming that the BPS configuration is localized
in the space and that the scalar field will fall off as
\be \phi \approx v' +  {\cal O}(\frac{1}{r^2})\ ,   \ee
where we choose the gauge where $v'$ is constant, one can determine the asymptotic behavior of $\Phi$ from
the Eqn. (\ref{electric1}) as
\be \label{leadingphi} \Phi =\frac{v }{g^2}  - \frac{Q_N}{8\pi^2  r^2 } + {\cal O}(\frac{1}{r^3})\ . \ee
The electric energy therefore becomes
\be Q_E  =2\int d^4 {\bf x} \ \partial_m \tr_N( \phi D_m \Phi) =  \tr(  v' Q_N)\ . \ee
Taking into account of the positive kinetic energy, we should consider the parameter region
$|v'|<< (\kappa g^2)^{-1}$ for the Coulomb branch so that
\ba v&=&v' + \frac{\kappa g^2}{2} (v')^2 -\frac{\kappa g^2 {\bf 1}_N}{2N} \tr_N (v')^2
~\simeq~ v' \ . \ea

Our BPS configurations can also carry the conserved angular momentum
\ba \label{Noe angular mom}
  J_{mn} &=& \int d^4{\bf x}\ (x_m P_n -x_n P_m )\nonumber \\
  &=& 2\int_{S^3_\infty}r^3dS^p\ \big(
  x_m {\rm tr}\left(\Phi F_{np}\right)-
  x_n {\rm tr}\left(\Phi F_{mp}\right)\big) + 4
  \int d^4{\bf x}\  {\rm tr}\left(\Phi F_{mn}\right)\ ,
\ea
where we used for the last equality the expression of the linear momentum density
\be P_m ~=~ \tr (F_{mp}\{ E_p, 1/g_{\rm eff}^2\}) ~=~ 2\partial_p \tr (
F_{mp}\Phi)\ . \ee
The second term in Eqn. (\ref{Noe angular mom}) can also be translated to the surface term as
\begin{eqnarray} \label{Noe angular mom2}
  \int d^4{\bf x}\ {\rm tr}(\Phi F_{mn})
  &=&\int_{S^3_\infty}r^3\left(dS^m\alpha_n-dS^n\alpha_m+
  \epsilon_{mnpq}dS^p\alpha_q
  \right)\ ,
\end{eqnarray}
once the one-form $\alpha=\alpha_m dx^m$ satisfy the differential equation
\be \label{alpha}
  \left( 1 + \ast_4 \right) d\alpha~=~\text{tr}\left( \Phi F \right)\ .
\ee
The solution of the above equation for the BPS configurations will be presented in the next subsection.

\subsection{BPS Configurations}

To find out the BPS configurations, we need to first solve the
antiselfdual equation with $\alpha=-1$,
\be F_{mn}+\tilde{F}_{mn}=0 \ee
The general solutions of the antiselfdual equations are given the
ADHM method. In a given instanton background, one needs to solve
the second equation (\ref{gauss2}). Fortunately, the general
solutions of the second equations can be also found in terms of the
ADHM data.

Let us first summarize the ADHM method briefly. For details, it is referred to \cite{Dorey:2002ik}. Starting with the
constant matrices $(\bar{q}_i)_{N, K}$ and $(a_m)_{K,K}$, one can construct a $(N+2K)\times 2K $ matrix
\be  \Delta ~=~ \left( \begin{array}{cc} \left( \bar{q}_1 \hspace{0.1cm} \bar{q}_2 \right),  & M_{m}e^{m} \end{array}
\right)
\ee
where $e_{m}=(\sigma_{a},i)$ and   $M_m=a_m + {\bf 1}_K x_m$. Note
that $\bar{e}_m e_n = \delta_{mn} +i\eta^a_{mn}\sigma^a$ where
selfdual t'Hooft tensor $\eta^a_{mn}$ satisfies, for instances, $\eta^3_{12}=\eta^3_{34}=1$.
We introduce $V_{N+2K,N}(x)$ such that
\be \Delta^\dagger V~=~0\ , \hspace{1cm} V^\dagger V~=~ {\bf 1}_N\ . \ee
The condition on the ADHM data requires that
\be \Delta^\dagger \Delta ~=~ f^{-1}(x) {\bf 1}_2 \ee
with an invertible $K\times K$ matrix $f(x)$. This implies the
ADHM condition
\be i\eta^a_{mn}[a_m,a_n] + \sigma^a_{ij} q_j\bar{q}_i~=~0\ , \ee
where $a=1,2,3$. The ADHM method leads to the hermitian gauge field to be
\be A_\mu ~=~ i V^\dagger \partial_\mu V\  \ee
for the anti-instantons.

As shown in appendix, the solution of Eqn. (\ref{gauss2}) for the traceless scalar field is given by
\be \label{scalarsol} \Phi(x) ~=~ \bar{V} \left( \begin{array}{cc} \frac{v}{g^2} & 0 \\ 0  &
(\frac{\varphi}{g^2} -\frac{\kappa}{2} f(x))\otimes {\bf 1}_2
\end{array}\right) V  -\frac{\kappa}{4N}( \partial_m^2\ln{\rm det}f ) {\bf 1}_N  \ee
where the constant matrices $v$ and $\varphi$ satisfy $\tr_N v + \tr_K \varphi =0$ and
an  inhomogeneous linear matrix equation,
\be \frac{8\pi^2}{g^2} \Big(  {\bf L}\varphi + q_i v\bar{q}_i  \Big)  - 8\pi^2\kappa  {\bf 1}_K ~=~ 0
\label{Gauss2}  \ee
where ${\bf L}\varphi = - [a_m,[a_m,\varphi]]
-\frac12\{ q_i\bar{q}_i ,\varphi\} $.

We now in turn express the physical charges of those
BPS objects, dyonic instantons in terms of the AHDM data. From the leading order behavior
of the scalar field $\Phi$ (\ref{scalarsol}) at infinity
 \be \Phi ~=~  \frac{v}{g^2}  + \frac{1}{8\pi^2 r^2} \Big( \frac{8\pi^2 }{g^2} \big(
   \bar{q}_i \varphi q_i- \frac{1}{2}\{ \bar{q}_i q_i, v\}   \big)
     -\frac{ n K}{N} \Big) + {\cal O}(\frac{1}{r^3})\ ,  \ee
together with Eqn. (\ref{leadingphi}), one can read off the following electric charge of the $N \times N$ matrix
\be\label{electric}
  Q_N~=\frac{8\pi^2}{g^2}  \Big( \frac{1}{2}\{ \bar{q}_i q_i,v \} -q_i\varphi\bar{q}_i \Big) +\frac{nK}{N}{\bf 1}_N\ .
\ee
One can show from Eqn. (\ref{Gauss2}) that the electric charge matrix $Q_N$ is traceless. We now in turn
consider the angular momentum of the dyonic instantons. As shown in \cite{KL07}, one can solve Eqn.
(\ref{alpha}) using the various identities in the ADHM method
\begin{equation}
  \alpha_m~=~\frac i4 \ {\rm tr}\left(\mathcal{J}(P \partial_m P-\partial_m P P )+
  4 [\varphi,a_m] f - \frac{\kappa}{6}\epsilon_{mnpq}
  \partial_{n}f^{-1}f\partial_{p}f^{-1}
  f\partial_qf^{-1}f\right),
\end{equation}
where $P=V V^\dagger$ and ${\cal J}=\frac{1}{g^2} \text{diag}\left( v_N, \varphi_K {\bf 1}_2 \right)$. For the BPS
configurations, the angular momentum (\ref{Noe angular mom},\ref{Noe angular mom2}) therefore takes the form
as
\begin{equation}\label{angular mom}
  J_{mn}=- \frac{8\pi^2}{g^2} i\left\{  \frac12
  {\rm tr}_k \Big(  q_i v \bar{q}_j
  (\bar\sigma_{mn})_{ji} \Big)+(1+\ast_4)
  {\rm tr}_k\left(\varphi[a_m,a_n]\frac{}{}\right)\right\}\
  .
\end{equation}

\subsection{A Single Dyonic $SU(N)$ Instanton}

To be concrete, let us consider the profile of a single dyonic instanton in
the $SU(N)$ gauge theory. The solution can be obtained either by using the general
solution presented in the previous section, or simply solving it directly in the single instanton case. While the
former is discussed in appendix, we follow the latter.

The $4N$-dimensional moduli space of an  instanton is composed of 4-parameters
for the center of the instanton position and $4N-4$ parameters for the internal degrees
of freedom. The internal part is a haper-K\"ahler cone over 
a tri-Sasakian space  $SU(N)/U(N-2)$, where this symmetric space $SU(N)/U(N-2 )$ is a  
 quaternionic space $
\frac{SU(N)}{SU(2)\times U(N-2)}$ with a $SU(2)$ fiber.  The moduli space is spanned by
moduli of the single $SU(2)$ 't Hooft instanton, plus additional
moduli which determines the way how $SU(2)$ is embedded in
$SU(N)$. In this section we work in the `regular' gauge, opposed
to the singular gauge which is convenient for the ADHM instantons.

The $SU(N)$ single instanton gauge field is given as
\begin{equation}\label{single}
  A_m=T^aA^a_m=
  T^a\bar{\eta}^a_{ mn}\frac{2  x_n}{r^2+\rho^2}\ ,
\end{equation}
where $T^a$ generates arbitrary $SU(2)$ sub-algebra of $SU(N)$ and
$r=\sqrt{x_lx_l}$. We can take it to be the following:
\begin{equation}
  T^a=U^\dag\left(\frac{1}{2}\sigma^a\right)U, \hspace{1cm} \left[T^a, T^b\right]=i\epsilon^{abc} T^c, \label{Tdef}
\end{equation}
where
\begin{equation}
U_{2.N}=\left(\begin{array}{cccc} e_{11} & e_{12} & \cdots & e_{1N} \\
e_{21} & e_{22} & \cdots & e_{2N} \end{array}\right)
 \equiv \left( \begin{array}{c} \bar{{\bf e}}_1 \\\bar{{\bf e}}_2 \end{array}\right) \ , \ \bar{{\bf e}}_i \cdot {\bf e}_j =
\delta_{ij}\ .
 \end{equation}
In addition to the $SU(2)$ algebra, the generators $T^a$ satisfy the anti-commutation relation
\be  \{T^a,T^b\} =  \frac{1}{2} \delta^{ab}
U^\dagger U =\frac{\delta^{ab}}{N}{\bf 1}_N- \frac{\delta^{ab}}{2} T\ ,
\ee
where
 \be T\equiv   \frac{2}{N}{\bf  1}_N- U^\dagger U \ . \ee
The mutually orthogonal complex unit vectors
${\bf e}_i$ contain $4N-4$ real components. Among these, the
$U(1)$ phase rotation ${\bf e}_i\rightarrow{\bf e}_ie^{i\alpha}$
does not affect $T^a$ or $A_m$, so we may regard $\alpha$ as a
`gauge' degree. Together with the remaining $4N-5$ gauge-invariant
degrees, the scalar $\lambda$ provides the desired $4N-4$ zero
modes apart from the 4 translations that we ignore here.

We are interested in the solution of the traceless part of Eqn.
(\ref{gauss2})  for $\Phi$. For a single instanton (\ref{single}), the field strength can be described as
\be F_{mn} ~=~ \frac{4\rho^2}{(r^2+\rho^2)^2} \bar{\eta}^a_{mn} T^a \ ,\ee
from which one can obtain
\be F_{mn}^2 = \frac{16\rho^4}{(r^2+\rho^2)^4}  4T^a T^a
  ~=~  \frac{16\rho^4}{(r^2+\rho^2)^4}
 (\frac{6}{N}{\bf 1}_N - 3T)\ .
\ee
Taking the ansatz for $\Phi$ field as $\Phi=F(r)\ T$, the equation for $\Phi$ can be simply reduced to
\begin{equation}
  \partial_m^2 F(r) ~=~ -   \frac{12\kappa \rho^4}{(r^2+\rho^2)^4}
   ~=~ \frac{\kappa}{8} (\partial^2)^2\left(\log(r^2+\rho^2) \right)\ .
\end{equation}
The solution for $F(r)$ is therefore given by
\begin{equation}
 F(r)= \frac{\kappa}{2}\frac{r^2+ 2\rho^2 }{(r^2+\rho^2)^2}\ ,
\end{equation}
in perfect agreement with that  obtained from the ADHM
method.

Now we would like to obtain the scalar $\phi=\phi^aT^a$ from the
above $\Phi$. Our claim is that $\phi$ is also spanned by
the generator $T$ only as $T^2 -\tr_N T^2/N \propto  T$.
 Therefore we
try $\phi= h(r)T$, and obtain the following equation for $h(r)$:
\begin{equation}
  \frac{1}{g^2}h(r)+\frac{\kappa}{2}\frac{4-N}{N}h(r)^2=F(r)\ .
\end{equation}
The solution  $h_N(r)$  for the $SU(N)$ gauge theory is
\begin{eqnarray}
  h_4(r)&=&g^2 F(r)\ \ \ ({\rm for}\ N=4)\\
  h_N(r)&=&-\frac{1}{\kappa g^2}\frac{N}{4-N}\pm
  \sqrt{\left(\frac{1}{\kappa g^2}\frac{N}{4-N}\right)^2+\frac{2}{\kappa}
  \frac{N}{4-N} F(r)} , \ \ (N\neq 4)
\end{eqnarray}
where the sign in front of the square-root for the second case is
fixed by requiring that $\phi=0$ at $r=\infty$.

To summarize, the scalar solution, and therefore the whole
instanton solution in the presence of Chern-Simons term is
classified into the following three cases:
\begin{enumerate}

\item For $SU(3)$, the scalar solution is given as
\begin{equation}
  h_3(r) = -\frac{3}{\kappa g^2}+\sqrt{\left(\frac{3}{\kappa g^2}\right)^2 +
    \frac{3r^2 + 6\rho^2}{(r^2+\rho^2)^2} }\ .
\end{equation}
The solution is well-defined for all $\rho>0$.

\item For $SU(4)$, the scalar is given as
\begin{equation}
    h_4(r)=\frac{\kappa g^2}{2}\frac{r^2+2\rho^2}{(r^2+\rho^2)^2}\ .
\end{equation}

\item For $SU(N)$ with $N\geq 5$, the solution is given as
\begin{equation}
   h_N(r)=\frac{1}{\kappa g^2}\frac{N}{N-4}-
  \sqrt{\left(\frac{1}{\kappa g^2}\frac{N}{N-4}\right)^2-
  \frac{N}{N-4}\frac{r^2+2\rho^2}{(r^2+\rho^2)^2}}\ .
\end{equation}
The solution is well-defined at all $r\geq 0$ only in the
background of large enough instanton, namely,
\begin{equation}\label{minimal scale}
  \rho \geq \rho_c\equiv \sqrt{\frac{2(N\!-\!4)}{N}}  \kappa g^2  \ .
\end{equation}
For $\rho$ smaller than this value $\rho_c$, there is no \textit{real
solution} $\phi(r)$ for $r<r_0$, where
\begin{equation}
4 r_0^2=   \rho_c^2-4\rho^2 + \sqrt{\rho_c^4+8\rho^4}  \ .
\end{equation}

\end{enumerate}

For large $\kappa$, there is however no meaning of the classical instanton of size of order $\kappa g^2= n g^2/8\pi^2 $,
which is quite bigger than the ultraviolet scale $g^2$.    Although the UV completion of theory, or the reliable UV description of those solutions, seems to be
important, we will not discuss about it any more.

\section{Low Energy  Dynamics }

\subsection{One-dimensional $U(K)$  Gauge Theory}

The 16 supersymmetric $SU(N)$ gauge theory arises from the low energy dynamics  of $N$ D4 branes. It is well-
known that the four-dimensional $K$ instanton solitons are realized as $K$ D0 branes on D4 branes. The low
energy dynamics of instantons can be described by the one-dimensional  $U(K)$ gauge theory with eight
supersymmetries. The bosonic part of the Lagrangian is
\begin{eqnarray}\label{LED}
  {\cal L}_0 &=
  \frac{M}{2}  &\tr_K \Big\{
  \left(D_0 a_m \right)^2  +   \left| D_0 q_i  \right|^2   + \left[\varphi,  a_m   \right]^2
  - \left|( \varphi q_i  - q_i v )\right|^2      \nonumber\\ &~&
  +c \left(D_0 \varphi \right)^2 + c \left(D^a\right)^2
  +  D^a \left(  \sigma^a_{ij}  q_j \bar{q}_i +i \eta^a_{mn} \left[a_m , a_n \right]
  \right)   -\zeta^a D^a  \Big\}\ ,
\end{eqnarray}
where $M=8\pi^2/g^2$ denotes the instanton mass parameters. This Lagrangian also has a $SU(N)$ flavor
symmetry, understood as the gauge symmetry in the five-dimensional gauge symmetries. Under the flavor
symmetry, the hypermultiplet scalars $q_i, i=1,2$ ($K\times N$ matrices) transform as
\be
  q_i \ \ \rightarrow \ \ q_i' ~=~ q_i U \hspace{2cm} (U \in SU(N))\ .
\ee
Another hypermultiplet scalars $a_m, m=1,2,3,4$ denote the matrix position of $K$ instantons
on $D4$ branes. There are two controllable parameters of the present model of which one is the $N\times N$
hermitian matrix $v$, representing the positions of $N$ D4-branes, and another is the Fayet-Iliopoulos parameter
$\zeta$, necessary for the noncommutative field theory. As well as the gauge symmetry $U(K)$ and flavor
symmetry $SU(N)$, the theory has $SO(4) \simeq SU(2)_R\times SU(2)$ global symmetry. For instances, $a_m$
furnish a vector representation under $SO(4)$ while $q_i$  are in a fundamental representation under $SU(2)_R
$. The arbitrary parameter $c$ of dimension {\it mass}$^4$ can be determined from this D0--D4 system by
$c=2\pi/(2\pi l_s)^4$.

In this work, our 5-dim theory has  8 supersymmetries and a Chern-Simons term. Instantons of our theory is again
1/2 BPS with now four supersymmetries and so  their property will be
different from those from  16 supersymmetric theories. One may wonder what the low energy dynamics of
instantons with four supersymmetries. As far as the classical feature is concerned, our analysis of the BPS
equations tells us that the gauge field configuration does not change. Only our instantons now carry some electric
charge due to the Chern-Simons term. Thus we expect a simple modification of the above Lagrangian may
suffice. Indeed, here we argue that it is sufficient to add to the Lagrangian (\ref{LED})  the following 1-dim
Chern-Simons term
\be \label{LCS}
{\cal L}_{cs} =  n\, \tr_K ( A_0 -\varphi ) ~.
\ee

Let us first look at the supersymmetry. Note that $a_m, q_i$ are hypermultiplet and $A_0,\varphi,$ are part of the
vector multiplet. In 10-dim notation for the vector multiplet, the above Chern-Simons term is invariant under
supersymmetric transformation
\be \delta (A_0-\varphi)=\bar{\lambda}(\Gamma_0-\Gamma_5)\epsilon =
 \bar{\lambda}\Gamma_0(1-\Gamma^{05})\epsilon
=0 \ee
if $\Gamma^{05}\epsilon= \epsilon$ . Thus imposing additional condition on the spinor parameter breaks the
supersymmetry to half. It is to be noted that the 1-dim Chern-Simons term introduces a background electric charge
to $U(K)$ theory. With the presence of the fundamental representation of $U(K)$, the invariance of the path
integral amplitude under a local $U(1)$ gauge transformation $e^{i2\pi \Lambda(t) T}$ such that $\Lambda(\infty)-
\Lambda(-\infty)=
2\pi l$  where $T= {\rm diag}(1,0,0,...0)$, leads to the quantization $n\in  {\bf Z}$.

We then consider the stringy origin of the one-dimensional Chern-Simons term (\ref{LCS}). From our D-brane
picture in section 2, the instantons were identified as D1-branes stretched between NS5-branes. The open strings
connecting flavor D7 or D5 branes to D1-instantons leads to the massive fundamental hypermultiplets in the
matrix model (\ref{LED}). Their one-loop effect indeed induces the very 1-dimensional Chern-Simons term.

The ground state should have the lowest energy. First of all the D-term equation implies that
\be
  D^a = \frac{1}{2c}  ( \zeta^a - i\bar{\eta}^a_{mn}[a_m,a_n] -\sigma^a_{ij} q_j \bar{q}_i )\ .
\ee
Since the Chern-Simons term does not affect the energy, the energy of the Lagrangian $L_0 + L_{cs}$ can be
rearranged as
\begin{eqnarray}
  {\cal E} &=& \frac{M}{2}   \tr_K\Big\{  \left(D_0 a_m + i \left[ \varphi, a_m
  \right]\right)^2  + \left|
  D_0 q_i + i \left( \varphi q_i - q_i v \right) \right|^2  + c \left(D_0 \varphi \right)^2 +  c \left( D^a \right)^2 \Big\}
\nonumber  \\
  &~&
  +M \tr_K  \Big\{ \varphi \Big( - i \left[a_m , D_0 a_m  \right]
  - \frac{i}{2} \big( q_i D_0 \bar{q}_i - D_0 q_i  \bar{q}_i \big)
  \Big) +\frac{n}{M}    \Big\}  \nonumber \\ &~& +  \frac{M}{2} \text{tr}_N \Big\{ i \left( D_0
  \bar{q}_i q_i - \bar{q}_i D_0 q_i \right) v \Big\}
\end{eqnarray}
After using the Gauss law constraint
\be M\Big( - i[a_m, D_0 a_m]-\frac{ i}{2}(q_i D_0\bar{q}_i - D_0q_i\bar{q}_i )  \Big) +    n{\bf 1}_K =0 \ ,
\label{gauss4} \ee
we can obtain a BPS bound on the energy
\be {\cal E} \ge \tr_N(vQ_N) \ , \label{bpsbd} \ee
where the $SU(N)$ flavor charge is given by
\be \label{flavorcharge} Q_N= M\Big(  \frac{ i}{2} (D_0\bar{q}_i q_i - \bar{q}_i D_0 q_i )  - \frac{ i}{2N}   \tr_N
(D_0\bar{q}_i q_i -
\bar{q}_i D_0
q_i ){\bf 1}_N
\Big)\ . \ee

The BPS configurations which saturate the bound should satisfy
\be \label{ADHMconstraint} cD^a = \zeta -\bar{\eta}^a_{mn}[a_m,a_n] - \sigma^a_{ij}q_j \bar{q}_i =0 \ee
which is nothing but the ADHM constraint. We started with $4K^2$-parameterized $a_m$ and $4NK$ parameterized
$q_i$. Imposing the $3K^2$ ADHM constraint and taking out $K^2$ $U(K)$ gauge degrees of freedom leads to
$4KN$ moduli-space coordinates $z_A, A=1,2,...,4KN$. The solutions of the above equation would be given by
$a_m(z_A)$ and $q_i(z_A)$. The rest of the BPS equations are
\be D_0 a_m +i[\varphi,a_m]=0,\ D_0q_i+i(\varphi q_i - q_i v)=0 , \ D_0\varphi=0 \ . \ee
In the gauge $A_0=\varphi$, the configuration is static in time, besides $q_i \sim q_{i0} e^{ivt}$. For the BPS
configurations, the Gauss law becomes
\be \label{gauss5} M\Big( - [a_m,[a_m,\varphi]]  - \frac{1}{2} \{ q_i\bar{q}_i, \varphi\}  + q_i v\bar{q}_i \Big)  -
8\pi^2\kappa   {\bf
1}_K =0 \ , \ee
which is identical to Eqn. (\ref{Gauss2}) in the field theory, once we noting that $8\pi^2\kappa = n$. This leads to a
further support for the validity of our low energy Lagrangian for the instanton solitons. As the kinetic part for $
\varphi$ is positive-definite, the solution would be a linear combination,
\be \varphi = \varphi_v(z_A)  + \varphi_\kappa (z_A) \ee
where ${\bf L}\varphi_v + q_i v \bar{q}_i=0$ and $M {\bf L} \varphi_\kappa  - 8\pi^2\kappa = 0 $.

Let us now in turn consider the various charges that the BPS configurations can carry. The traceless $SU(N)$
flavor charge (\ref{flavorcharge}) becomes
\be\label{electric2} Q_N= -M\Big( \bar{q}_i \varphi q_i -\frac{1}{2} \{ \bar{q}_i q_i, v\}\Big)  +\frac{nK }{N}{\bf
1}_N  \ , \ee
which is in the agreement with the electric charge (\ref{electric}) of the dyonic instantons. We then compute the
angular momentum of this half BPS solution
\begin{eqnarray}
  J_{mn} &=&-i M \text{tr}\left( \frac14 \left( q_i D_0
  \bar{q}_j - D_0q_i \bar{q}_j \right)
  \left(\bar{\sigma}_{mn}\right)_{j i}
  + \left( D_0 a_m a_n - a_m D_0a_n \right) \right)
  \nonumber \\ &=&-iM \text{tr} \left( \frac12 \left( q_i v \bar{q}_j -
  \varphi q_i \bar{q}_j \right) \left(\bar{\sigma}_{mn}\right)_{ji}+
  2 \varphi \left[a_m,a_n \right] \right) \nonumber \\
  &=& -iM \text{tr} \left( \frac12 \left( q_i v \bar{q}_j \right)
  \left(\bar{\sigma}_{mn}\right)_{ji} + \left(1 + \ast_4
  \right) \left( \varphi \left[a_m,a_n \right] \right)
  \right)\ ,
\end{eqnarray}
where for the last equality we used the ADHM constraints (\ref{ADHMconstraint}). This angular momentum agrees
again with the result (\ref{angular mom}) in field theory analysis. It is therefore conceivable to identify the dyonic
instantons with the BPS configurations of the present 1-dim gauge theory.

\subsection{Moduli Space Dynamics}

The vacuum configuration modulo the gauge equivalence is the moduli
space. The low energy dynamics is described by the nonlinear sigma model
of the moduli coordinates.  For the $K$ instantons of the $SU(N)$ gauge theory, the nonlinear sigma model is characterized by the $4KN$-dimensional hyper-K\"ahler space,  with several Killing vectors for symmetry of the instanton moduli space.  The Coulomb phase leads
the natural potential on the moduli space. We argue that the 1-dim Chern-Simons term
leads to the magnetic field on the instanton moduli space.
We may wonder how the Chern-Simons effect can be
realized in the moduli space dynamics.  In terms of the moduli space coordinates $z^A$, the dynamics can be described by the
Lagrangian
\be L ~=~ \frac{M}{2} g_{AB}(z)  \dot{z}^A \dot{z}^B  + n\, \dot{z}^A {\cal A}_A(z) - U(z) \ .\ee
(A similar magnetic field  has appeared in the moduli space  of
magnetic flux vortices in the Abelian Higgs model with a background charge and a Chern-Simons terms~\cite{Kim:2002qma}.)

The metric for the moduli space can be determined by the standard procedure. We start with ignoring the effect of
the Chern-Simons term $\kappa$ and the vacuum parameter $v$. One way to
approach is to put the initial field configuration,
\be a_m(z), \   q_i(z),  \ee
determined by the moduli coordinate, and its time derivative
\ba &&
 \hat{D}_0 a_m ~=~ \partial_0a_m -i[\hat{A}_0,a_m] ~=~  \dot{z}^A\hat{D}_A a_m
  ~=~ \dot{z}^A \Big(  \frac{ \partial a_m}{\partial z^A}  -i [\Lambda_A, a_m] \Big) ,
 \nonumber \\
 &&
 \hat{D}_0q_i ~=~ \partial_0 q_i -i \hat{A}_0 q_i ~=~ \dot{z}^A\hat{D}_Aq_i
 ~=~ \dot{z}^A \Big( \frac{\partial q_i}{\partial z^A } -i(\Lambda_A q_i -q_i  \Big)\ ,
\ea
which is given by the velocity of the moduli coordinate. The initial configuration should satisfies the Gauss law
constraint (\ref{gauss4}), implying that
\be {\bf L}\hat{A}_0 ~=~ -i[a_m \partial_0 a_m]-\frac{i}{2}(q_i\partial_0\bar{q}_i -\partial_0 q_i \bar{q}_i )  \ . \label{ah0eq} \ee
Inserting this initial field configuration back to the $U(K)$ gauge theory Lagrangian (\ref{LED}), one can obtain the
metric on the moduli space
\be \label{metric} g_{AB}(z) ~=~ \tr_K \Big( \hat{D}_A a_m \hat{D}_B a_m
+ \frac{1}{2} (\hat{D}_A q_i \hat{D}_B \bar{q}_i + \hat{D}_B q_i \hat{D}_A \bar{q}_i ) \Big)\ .
\ee

With the Chern-Simons term and the vacuum expectation value $v$, we modify the field velocity of
the previous initial field configurations to be
\be D_0 a_m ~=~ \hat{D}_0 a_m -i [\Delta A_0, a_m ],\   D_0 q_i ~=~ \hat{D}_0 q_i -i \Delta A_0 q_i   \ee
and suppose the scalar field $\varphi $ relaxes to the lowest
energy configuration $\varphi=\varphi_\kappa +\varphi_v$ as before.
The Gauss law and the scalar field equation tells us $\Delta A_0 =  \varphi_\kappa$ and
the $U(K)$  gauge theory Lagrangian now can take the form as
\be L ~=~ \frac{M}{2} g_{AB} \dot{z}^A \dot{z}^B
+ 8\pi^2 \kappa (  \hat{A}_0 -\varphi_v) -U(z)  \ , \label{last} \ee
from which one can find various corrections to the instanton moduli space dynamics.
Defining the gauge field in the moduli space to be
\be \label{oneform} {\cal A}_A (z) ~=~ \tr_K  {\bf L}^{-1}\Big(   -i \big[a_m, \frac{ \partial a_m}{\partial  z_A}  \big]
 -\frac{i}{2} \Big( q_i\frac{\partial
\bar{q_i}}{\partial z^A}  -\frac{\partial q_i }{\partial z^A} \bar{q}_i \Big) \Big) \ ,  \ee
the Chern-Simons term becomes
\be 8\pi^2\kappa \  \tr_K(\hat{A}_0) ~=~ 8\pi^2\kappa \  \sum_A \dot{z}^A {\cal A}_A(z)  \ .  \ee
Finally the potential term
\be U(z) ~=~ \frac{M}{2} \tr_K \Big(- [a_m, \varphi_v ]^2 + (\varphi_v  q_i -q_i v)
 (\bar{q}_i \varphi_v - v\bar{q}_i ) \Big) = \frac{M}{2}\tr_K \Big (-\varphi_v q_iv\bar{q}_i + q_i v^2 \bar{q}_i \Big)
\ee
is of order ${\cal O}(v^2)$ correction.  As well as ${\cal O}(\kappa)
$ correction $8\pi^2 \kappa \hat{A}_0$, there is of ${\cal O}(\kappa v)$ correction $- 8\pi^2 \kappa \varphi_v$.

We can find the Killing vector related to the $SU(N)$ gauge symmetry and the potential more explicitly. We choose the Higgs vacuum expectation value $v$ to be diagonal. The unbroken $U(1)^{N-1}$ generators are $N-1$ traceless matrices $T^a$ such that $\tr T^a T^b=\delta^{ab}/2$ and $v=v^a T^a $.  The $U(1)^{N-1}$ phase rotation (4.2) of the matter part  $q_i$ is $U=e^{i \zeta^a T^a}$, which  leads to the cyclic coordinates $\zeta^a$ and the    $N-1$  commuting Killing vectors, 
\be K_a^A\frac{\partial}{\partial z^A} =\frac{\partial}{\partial \zeta^a} .  \ee
With some effort similar to the discussions in Ref.~\cite{piljin},  one can show that  the moduli space dynamics (\ref{last}) becomes
\be L=\frac{M}{2} g_{AB} \dot{z}^A\dot{z}^B+ 8\pi^2\kappa (\dot{z}^A- G^A){\cal A}_A - \frac{M}{2} g_{AB}G^A G^B , \label{mosplag} \ee
where $G^A= K^A_a v^a $.  
This moduli space dynamics is of order $\dot{z}^2, \kappa\dot{z}, \kappa v, v^2$.    
We have ignored other higher order corrections. Calling the rest of the relative  moduli space coordinates $y^i$, we can re-express the above low energy
dynamics as
\ba L &=& \frac{M}{2}  h_{ij}(y)\dot{y}^i \dot{y}^j+ \frac{M}{2} k_{ab}(y)(\dot{\zeta}^a+\dot{y}^i\omega^a_i(y))(\dot{\zeta}^b+\dot{y}^j\omega_j^b(y)) \nonumber  \\
& & + 8\pi^2\kappa \ \dot{y}^i {\cal C}_i(y) + 8\pi^2\kappa\  (\dot{\zeta}^a-v^a){\cal V}_a(y) -\frac{M}{2} k_{ab}(y) v^a v^b 
\ea
with appropriate redefinition of the various terms. 

Now one can easily give a BPS bound on the energy related to the above Lagrangian and can obtain
the ground state energy as
\be E_{ground} =  v^a p_a  \ee
where $p_a$ is the canonical momentum related to the cyclic coordinates. This ground state energy is equivalent to the minimum of the bound (\ref{bpsbd}).     The BPS ground state configuration satisfies
\be \dot{y}^i=0\, , \ \dot{\zeta}^a = v^a  \ee
It would be interesting to find the supersymmetric extension of the moduli space dynamics (\ref{mosplag}).

\subsection{A Single Chern-Simons Instanton Moduli Dynamics}

Let us consider the moduli space dynamics of a single instanton. As discussed before, its moduli space is a $4N $-dimensional hyper-K\"{a}hler manifold. For generality, we will turn on Chern-Simons level $\kappa$ and
noncommutative parameter $\zeta$, but will stay in the symmetric phase   $v=0$.  It implies that we will analyze the moduli space of a single instanton in the $U(N)$
gauge theory on noncommutative $R^4$ space rather than commutative space. We can of course take the
commutative limit to the $SU(N)$ gauge theory instanton by letting $\zeta \rightarrow 0$ together with decoupling
$U(1)$ factor. The single instanton moduli space for $U(N)$ gauge group is known to be the Calabi space
of dim 4N which is complete nonsingular hyper-K\"ahler space with   cohomogeneity one.

Before the detailed analysis, we first summarize the conventions used in Ref.~\cite{CGLP01} to
find the moduli space metric. For the invariant one-forms $L^i_{\ j}$ of $SU(N)$($\subset U(N)$) defined as
\begin{eqnarray}
  U d U^\dagger ~=~ - i L^i_{\ j} t_i^{\ j}\ , \ \ \ \ \  [
  t_i^{\ j}, t_k^{\ l} ] ~=~ \delta_k^{\ j} t_i^{ \ l} -
  \delta_i^{\ l} t_k^{\ j}~,
\end{eqnarray}
with $( t_i^{\ j} )^\dagger = t_j^{\ i}$ (or $(
L^i_{\ j} )^\dagger = L^j_{\ i}$) and $L^i_{\ i}=0$, the
Maurer-Cartan equations are
\begin{eqnarray}
  d L^i_{\ j}~=~ i L^i_{\ k} \wedge L^k_{\ j}~.
\end{eqnarray}
Here $i$ run over $\left(1,2, a \right)$, where $a$ denote the $SU(N-2)$ indices.
One can then define the generators of the
coset group $SU(N)/\left(SU(N-2)\times U(1) \right)$ as
\begin{eqnarray}
  \sigma_a~=~L^1_{\ a}\ , \ \ \ \Sigma_a~=~L^2_{\ a}\ , \ \ \
  \mu~=~L^1_{\ 1} - L^2_{\ 2}\ , \ \ \ \nu~=~L^1_{\ 2}~,
\end{eqnarray}
and those of $SU(N-2) \times U(1)$ as
\begin{eqnarray}
  \tilde{L}^a_{\ b}~=~ L^a_{\ b} + \frac1n Q \delta^a_{\ b}\ , \ \ \
  Q~=~ L^1_{\ 1} + L^2_{\ 2}~.
\end{eqnarray}
With those new basis, the Maurer-Cartan equations of our interests, for examples,
take the form as
\begin{eqnarray}
  d \mu &=& 2i \nu \wedge \bar{\nu} + i \sigma_a \wedge \bar{\sigma}^a
  - i \Sigma_a \wedge \bar{\Sigma}^a\ , \ \ \ dQ ~=~ i \sigma_a \wedge \bar{\sigma}^a
  + i \Sigma_a \wedge \bar{\Sigma}^a~.
\end{eqnarray}

Let us first consider the D-term conditions. Since for a single instanton $a_m$ are simple c-numbers, we can
solve explicitly the D-term condition with $\zeta^a=\delta^{a3}\zeta^2 $ as
\be \left(\begin{array}{c} q_1 \\  q_2\end{array}\right) = \frac{1}{\sqrt{2}} \Big( \begin{array}{cccc} \sqrt{u^2 +
\zeta^2} & 0 & \cdots & 0 \\
  0 & \sqrt{u^2-\zeta^2}  & \cdots & 0 \end{array}\Big) U  \equiv q_0 U
  \ee
where $U$ is an $U(N)$ element. The two useful quantities in our discussion are summarized as
\begin{eqnarray}
  && \hspace{-0.5cm} \text{tr} \Big\{ \left( \bar{q}_0 q_0 \right) U d U^\dagger
  \Big\} = -\frac{i}{2} \left( \zeta^2 \mu + u^2 Q \right) - id\alpha\\
  && \hspace{-0.5cm} \text{tr} \Big\{ \left(\bar{q}_0 q_0 \right)
   U d U^\dagger~ U d U^\dagger \Big\} = - \left(
   u^2  \left( \frac{Q^2 + \mu^2}{4}
  + \nu \bar{\nu} \right) + \frac{\zeta^2}{2} Q \mu +\frac{u^2+\zeta^2}{2}
  \sigma_a \bar{\sigma}^a +\frac{u^2-\zeta^2}{2}
  \Sigma_a \bar{\Sigma}^a \right),  \nonumber
\end{eqnarray}
where $d\alpha$ denotes the one-form corresponding to $U(1)$($\subset U(N)$) generator that rotates overall
phase of $q_i$. For the single instanton, the moduli space metric (\ref{metric}) of $U(N)$ non-commutative theory
can be described as
\begin{eqnarray}
  \hspace{-0.5cm}
  ds^2=\left( 1 - \frac{\zeta^4}{u^4} \right)^{-1} du^2
  + \frac{u^2}{4} \left( 1 - \frac{\zeta^4}{u^4} \right) \mu^2
  + u^2 \nu \bar{\nu} + \frac{u^2+ \zeta^2}{2}
  \sigma_a \bar{\sigma}^a + \frac{u^2-\zeta^2}{2}
  \Sigma_a \bar{\Sigma}^a,
\end{eqnarray}
which is known to be Calabi metric  \cite{CGLP01}. The
one-form gauge field (\ref{oneform}) on the moduli space now
becomes
\begin{eqnarray}
 {\cal A} _A dz^A   ~=~     \frac{\zeta^2}{2 u^2} \mu +
\frac{1}{2}   Q   + d\alpha~,
\end{eqnarray}
whose field strength is therefore
\begin{eqnarray}
  F~=d{\cal A} ~=~\frac{ 1}{ 2u^2} \left( -2 \zeta^2 \frac{du}{u}
  \wedge \mu + 2 i \zeta^2 \nu \wedge \bar{\nu} + i\left( u^2 +\zeta^2
  \right) \sigma_a \wedge \bar{\sigma}^a + i\left( u^2 - \zeta^2
  \right) \Sigma_a \wedge \bar{\Sigma}^a \right). \nonumber \\
  \label{fstn}
\end{eqnarray}
As mentioned before, the single Chern-Simons instanton can carry the electric charge (\ref{electric2}) even in the
symmetric phase $v=0$
\begin{eqnarray}\label{elcharge3}
  Q_N~=~-  8\pi^2 \kappa  \left[ \frac{1}{N} \begin{pmatrix} \frac{N-2}{2}{\bf
  1}_2 & 0 \\ 0 & - {\bf 1}_{N-2} \end{pmatrix} - \frac{\zeta^2}{2u^2} \begin{pmatrix} \sigma_3 & 0 \\ 0 & 0
\end{pmatrix}\right],
\end{eqnarray}
and no angular momentum. We note that, for commutative $SU(2)$ gauge theory, the charge matrix $Q_N$
identically vanishes. It agrees with the fact that there is no non-abelian Chern-Simons coupling for $SU(2)$
group.

Especially for a single instanton in $U(2)$ gauge theory on noncommutative space,
the moduli space metric is Eguchi-Hanson space with the metric
the moduli space metric of the non-commutative $U(2)$ single
instanton now becomes
\begin{eqnarray}
  ds^2&=& \left(1-\frac{\zeta ^4}{u^4}\right)^{-1} du^2 + \frac{u^2}{4}
  \left( \sigma_1^{\ 2} + \sigma_2^{\ 2} \right) + \frac{u^2}{4} \left( 1 -
  \frac{\zeta^4}{u^4} \right) \sigma_3^{\ 2}~,
\end{eqnarray}
which is nothing but the Eguchi-Hanson metric. Here, $\sigma_i$ are the standard $SU(2)$ invariant one-forms
\begin{eqnarray}
  \sigma_1 &=& - \sin\psi d\theta + \cos\psi \sin\theta d\varphi
  \nonumber \\
  \sigma_2 &=& \cos\psi d\theta + \sin\psi \sin\theta d\varphi
  \nonumber \\
  \sigma_3 &=& d\psi + \cos\theta d\varphi~,
\end{eqnarray}
which satisfy $d\sigma_i = \frac12 \epsilon_{ijk} \sigma_j \wedge
\sigma_k$. The gauge field $ {\cal A}_A $ can be
expressed as
\be
  {\cal A} =~   \frac{\zeta^2}{2 u^2} \sigma_3 +
  d\alpha~,
\ee
whose field strength now becomes the unique normalizable self-dual
harmonic two-form on the Eguchi-Hanson space\cite{CGLP01}
\begin{eqnarray}
  {\cal F}~=~d{\cal A}  ~=~ \frac{\zeta^2}{2u^2} \left( -\frac{2}{u} du \wedge
  \sigma_3 + \sigma_1 \wedge \sigma_2 \right)\ .
\end{eqnarray}
When we take the commutative limit $\zeta \rightarrow 0$ with ignoring overall $U(1)$ factor, the moduli space is
reduced to the orbifold $C^2/Z_2$ and the gauge field $A_A$ disappears. This disappearance is again
consistent to the non-existence of the single Chern-Simons instanton for $SU(2)$. For two identical instantons in the $U(1)$ theory on noncommutative space, the moduli-space metric is again 
Eguchi-Hanson and the 1-form ${\cal A}$ is again identical to that of the single instanton in the $U(2)$ theory as
shown in the previous paragraphs~\cite{LTY00}.

For $U(N)$ with $N>2$,  one can easily see that the $(N-1)$ wedge product of the magnetic field strength (\ref{fstn})  is  of order one in the radial coordinate $u$, making  it non-normalizable. Thus this product is not
the normalizable middle form discussed in Ref.~\cite{CGLP01}.  One  concludes that the Eguchi-Hanson   case is an exceptional one.

\subsection{Quantization of A Single Instanton Moduli Space}

We now restrict our attention only on the commutative five-dimensional field theories, i.e., $\zeta \rightarrow 0$ as
well as ignoring $d\alpha$ ($U(1)$) effect. Since the asymptotic Higgs vev is not developed, non-abelian gauge
symmetry $SU(N)$ is not broken. We therefore expect that a single Chern-Simons instanton can carry charges in
a certain representation under the gauge symmetry $SU(N)$. In order to specify the representation, we will
discuss the quantization
of the single Chern-Simons instanton moduli space and then find out the ground state degeneracies.

In the $\zeta \rightarrow 0$ limit, the metric of the moduli space becomes
\begin{eqnarray}
  ds^2~=~du^2+ \frac{u^2}{4} \mu^2
  + u^2 \nu \bar{\nu} + \frac{u^2}{2}
  \sigma_a \bar{\sigma}^a + \frac{u^2}{2}
  \Sigma_a \bar{\Sigma}^a ~,
\end{eqnarray}
while the Chern-Simons effect on the moduli space for the present model is given by
\begin{eqnarray}\label{u(1)gauge}
  8\pi^2 \kappa{\cal A}_A dz^A ~=~ \frac{8 \pi^2 \kappa}{2} Q\ .
\end{eqnarray}
One can show that the uniform magnetic flux $F$, proportional to the K\"{a}hler two-form on the coset space
${\cal G}=G/H\equiv\frac{SU(N)}{SU(2)\times SU(N-2)\times U(1)}$, is now turned on in the moduli space
\begin{eqnarray}
 8\pi^2\kappa  {\cal  F}~=~  \frac{ 8\pi^2  i \kappa  }{2}  \left( \sigma_a \wedge \bar{\sigma}^a + \Sigma_a \wedge \bar{\Sigma}^a
\right) \ , \label{magnetic}
\end{eqnarray}
from which the quantization of the Chern-Simons coefficient
\begin{eqnarray}
  8\pi^2 \kappa~=~ n\ ,
\end{eqnarray}
could be  also understood as the Dirac quantization.  While we have not worked out the detail, one can see that the coset space becomes $\mathbf{CP}^2$  when $N=3$, which has non-contractable two-cycles. Thus, we might assume the existence of two non-contractable two-cycles on the coset space for $N\ge 3$. The Dirac quantization could be understood as the quantization of the total flux on each such two-cycle. 

As we noted before, the moduli space of a single instanton is a hyper-K\"ahler cone over the tri-Sasakian space
$SU(N)/ U(N-2)$, which is a $SU(2)$-bundle over the coset space ${\cal G}=SU(N)/SU(2)\times SU(N-2)\times U(1) $.  Thus, the moduli space  consists  of the radial motion along the cone direction, a $SU(2)$ motion and the coset space motion. We are interested in the ground state dynamics on the moduli space with magnetic field only along the coset
space as one can note from Eq.~(\ref{magnetic}). Thus, the radial motion would be irrelevant  in the discussion and the wave function along the  $SU(2)$-fibre would settle down as the s-wave on $S^3$.    The problem in considerations would be reduced to  
the Landau problem of a charged particle moving on the coset space ${\cal G}$ under the background
$U(1)$ gauge field. Somewhat similar problem has been  analyzed extensively in \cite{KN02,KN06}. As a shortcut, we will just sketch the basic
idea how to analyze the representation, or degeneracy of the ground states. For details, it is referred to
\cite{KN02,KN06}.

When a particle moves on the coset space ${\cal G}=G/H$ (with the background gauge fields), energy
eigenstates of the system should be in certain representations under $G$. Since the background gauge field
${\cal A}$ (\ref{u(1)gauge}) is turned on only along the $U(1)$ direction in $H$, the eigenstates are singlet
under $SU(2)\times SU(N-2)$, implying that they can be represented as ${\tiny \yng(1,1)}$, its complex
conjugate
${\tiny \overline{ \yng(1,1)}}$ and their symmetric tensor
products without any trace part. We will denote such representations by $({\bf p,q})$ ($p$-times symmetric
products of ${\tiny \yng(1,1)}$ and $q$-times products of {\tiny
$\overline{\yng(1,1)}$}). One can show from Eqn. (\ref{u(1)gauge}) that a state in $({\bf p,q})$ carries $U(1)$
charge $p-q$, which has to be identified as the background charge $8\pi^2 \kappa=n$, that is
\begin{eqnarray}
  p-q~=~n .
\end{eqnarray}

Assuming $n > 0$ without loss of generality, one can conclude that
$q$ can play the role of the Landau level index because higher dimensional representation states are in the
higher energy levels. For the lowest Landau level ($q=0$, or $p=n$), the degenerate ground states are
therefore in the representation of $\kappa$-times symmetric products of ${\tiny
\yng(1,1)}$, i.e.,
\def\Dots{\cdot\cdot}
\begin{equation}\label{representation}
  \underbrace{\Yboxdim20pt \yng(7,7)}_{n}\kern -2.5pt
  \ \ \ \ \  \left( \text{ $SU(N)$ Young tableaux } \right)\ .
\end{equation}

As non-trivial cross checks of these results, let us consider  the case of $N=2$. Since there is no Chern-
Simons terms for the gauge group $SU(2)$ in the five-dimensional theories, the instanton should be in a
singlet whatever value the coefficient $\kappa$ is. The result above in (\ref{representation}) agrees with this
expectation. For the case of $N=3$, the coset space can be identified as the projective space CP$^2$
whose lowest Landau levels under the n-unit magnetic field are well-known to be
\def\Dots{\cdot\cdot}
\begin{equation}
  \underbrace{ \Yboxdim20pt \yng(7)}_{n}\kern -2.5pt \ \ \ \ \
  \left( \text{ $SU(3)$ Young tableaux } \right)\ ,
\end{equation}
nothing but the complex conjugate of the representation above. One can therefore again see the agreement.

\vspace{1cm}

\noindent{\bf Acknowledgment}

 We would like to thank Sangmin Lee, Ho-Ung Yee, Piljin Yi and David Tong  for valuable discussions. This work is supported in
part by the KOSEF SRC Program through CQUeST at Sogang University
(KML), KRF Grants No. KRF-2005-070-C00030 (KML), and the KRF
National Scholar program (SL,KML).

\appendix

\section{the ADHM Formalism and the Higgs Solutions}

Using the ADHM method, the field strength can be expressed as
\be  F_{mn} =   -2  \bar{\eta}^a_{mn} \bar{V} b f \sigma^a b^\dagger
V \ , \ee
where $b^\dagger = (0_{2K,N}, {\bf 1}_{2K})$. As $\bar{\eta}_{mn}^a\bar{\eta}_{mn}^b=4\delta^{ab}$, we get
\ba \frac{F_{mn}^2}{16}  &=&  \bar{V}bf\sigma^a b^\dagger V\bar{V} bf\sigma^a b^\dagger V \nonumber \\
&=& 3 \bar{V}bf^2b^\dagger V - 3 \bar{V}b fM_m fM_mf b^\dagger V  +
i\bar{\eta}^a_{mn} \, \bar{V} bf \sigma^a M_m f M_n f b^\dagger V\ea
Note that
\be \partial_m f = -f(\partial_m f^{-1}) f = -2f  M_m f \ee
and
\be D_m(\bar{V}bfb^\dagger V) = -4\bar{V}bfM_m fb^\dagger V \ee
Thus
\be D^2_m(\bar{V}bfb^\dagger V) = 8 \bar{V}bf^2 b^\dagger V
- \frac{1}{2} F_{mn}^2 \label{f2eq} \ee
Thus, the hermitian field strength satisfies
\be \tr_N (F_{mn}^2) = \partial_m^2 \partial_n^2 \ln {\rm det}(f) \ee

 We present an ADHM
solution for the equation (\ref{gauss2})
\begin{equation}
  \mathcal{D}^2\Phi=\frac{\kappa}{4}F_{mn}^2\
\end{equation}
where fields are in $N\times N$ matrix representation. We now
replace $F^2$ by using Eq.~(\ref{f2eq}). This leads to
\be D_m^2\Phi=\frac{\kappa}{2}\Big( 8 \bar{V}bf^2 b^\dagger V
-D^2_m(\bar{V}bfb^\dagger V) \Big) \ee
or
\be D_m^2( \Phi +\frac{\kappa}{2} \bar{V}bfb^\dagger V ) =
4\kappa\bar{V}f^2b^\dagger V\ee
For simplicity we introduce
\be \Psi = \Phi   +\frac{\kappa}{2}\bar{V} bf b^\dagger V \ee
which satisfy
\be D_m^2\Psi = 4\kappa \bar{V}bf^2 b^\dagger V \ee
We solve this equation with the ansatz
\be \Psi =\frac{1}{g^2}  \bar{V}\left(\begin{array}{cc}  v & 0 \\
0 & \varphi \otimes {\bf 1}_2  \end{array}\right) V \ee
\be D_m^2 \Psi = \frac{4}{g^2} \bar{V}bf \Big( {\bf L}\varphi
+ q_i v\bar{q}_i   \Big) f b^\dagger V  \ee
where  a linear operator ${\bf L}$ is defined so that
\be {\bf L} \varphi\equiv - [a_m,[a_m,\varphi]]-\frac{1}{2}\{
q_i\bar{q}_i, \varphi\} \ee
Thus the equation $D_m^2\Psi=2\kappa \bar{V}bf^2b^\dagger V$ can be solved if
\be \frac{8\pi^2}{g^2}\Big(  {\bf L}\varphi+ q_i v \bar{q}_i \Big) - 8\pi^2\kappa  {\bf 1}_K =0 \ee
Note that the above equation is invariant under $(q,\varphi)\rightarrow (q,\varphi)+ c ({\bf 1}_N, {\bf 1}_K)$.
Once the solution is known, the traceless scalar field configuration would be
\be \Phi(x) =\bar{V}\left(\begin{array}{cc} \frac{v}{g^2} & 0 \\
0 & (\frac{\varphi}{g^2} - \frac{\kappa}{2}f(x))\otimes {\bf 1 }_2
\end{array}\right) V -\frac{\kappa}{4N}\partial^2_m \ln  {\rm det} f  \ee
where the traceless condition implies that   $\tr_N q + \tr_K \varphi=0$.

\end{document}